\documentstyle[12pt,a41,epsfig]{article}
\newcommand{\bq}{\begin{equation}}
\newcommand{\eq}{\end{equation}}

\newcommand\GeV{\,\mbox{GeV}}
\newcommand\TeV{\,\mbox{TeV}}

\newcommand\ka{\mbox{\boldmath $\kappa_a$}}

\begin{document}
\sloppy
\thispagestyle{empty}
\begin{flushleft}
DESY 96--219 \\
{\tt hep-ph/9610506}\\
October  1996
\end{flushleft}

\mbox{}
\vspace*{\fill}
\begin{center}
{\LARGE\bf Vector Leptoquark Pair Production in}\\

\vspace{1mm}
{\LARGE\bf  $e^+e^-$ Annihilation}\\

\vspace{2em}
\large
Johannes Bl\"umlein$^1$, Edward Boos$^{1,2}$, and
Alexander Kryukov$^{1,2}$
\\
\vspace{2em}
{\it $^1$DESY --Zeuthen,}
 \\
{\it Platanenallee 6,
D--15735 Zeuthen, Germany}\\
{\it $^2$Institute of Nuclear Physics, Moscow State University,}\\
{\it RU--119899 Moscow, Russia}\\
\end{center}
\vspace*{\fill}
\begin{abstract}
\noindent
The cross section for vector leptoquark pair production in $e^+e^-$
annihilation is calculated for the case of finite  anomalous gauge boson
couplings $\kappa_{\gamma, Z}$ and $\lambda_{\gamma, Z}$. The
minimal cross section is found to behave $\propto \beta^7$,
leading to weaker mass bounds in the threshold range than in models 
studied previously.
\end{abstract}
\vspace*{\fill}
\newpage

\vspace{2mm}
\noindent
Various extensions of the Standard Model predict bosonic states
carrying both lepton and quark quantum numbers, the leptoquarks.
If their couplings
are baryon and lepton number conserving, cf. see~\cite{BRW}, these
states may even exist in the mass range accessible at
high energy colliders.
Stringent bounds on the leptoquark--fermion couplings $\lambda_{lq}$ are
derived from low energy data~\cite{LEUR}. In the mass range
$M_V \leq 1\TeV$ they are required to be much
smaller than the
electromagnetic coupling.
On the other hand
the bosonic leptoquark couplings cannot be arbitrarily small but are
determined by the respective gauge couplings and eventual anomalous
couplings.

In this note we extend the analysis performed in a previous
investigation~\cite{BR}, where minimal vector
couplings~\cite{BD}  $\kappa = 1$ and $\lambda = 0$ were considered, to
the general case of four independent anomalous couplings
$\kappa_{\gamma},  \kappa_{Z},  \lambda_{\gamma}$ and $\lambda_Z$
for the process of vector leptoquark pair production in $e^+e^-$
annihilation, $e^+e^- \rightarrow \overline{V} V$~\footnote{
Since the leptoquark states of ref.~\cite{BRW}
were constructed demanding $SU(2) \times U(1)$ symmetry 
it appears natural to us to consider {\it both} the anomalous
couplings $\kappa_{\gamma,Z}$ and $\lambda_{\gamma,Z}$
in the effective Lagrangian,
following~\cite{KALAX}.}. Due to the smallness of the fermionic couplings
we will consider the bosonic contributions only in the following.
The annihilation
 process dominates at not too large cms energies. Additional
contributions are due to the photon--photon fusion processes,
$e^+e^- \rightarrow e^+e^- \overline{V} V X$, which were studied
in refs.~\cite{BB, BBK}.

The tri-linear coupling of the photon and $Z$-boson to a pair of
vector leptoquarks is given by
\begin{equation}
\label{eqVERT}
V_{\mu_{1} \mu_{2} \mu_{3}}^{\overline{V} V \gamma (Z)}
(k_1, k_2, k_3)  =
e Q_{\gamma (Z)}(V)
 \bigg[ \widehat{V}_{\mu_{1} \mu_{2} \mu_{3}} +
\kappa_{\gamma(Z)} \widehat{V}_{\mu_{1} \mu_{2} \mu_{3}}^{\kappa}
+ \frac{\lambda_{\gamma(Z)}}{M_V^2} \widehat{V}_{\mu_1 \mu_2 \mu_3}
^{\lambda} \bigg],
\end{equation}
where
\begin{eqnarray}
\widehat{V}_{\mu_1, \mu_2, \mu_3}(k_1, k_2, k_3) &=&
  (k_1 - k_2)_{\mu_3} g_{\mu_1 \mu_2}
+ (k_2 - k_3)_{\mu_1} g_{\mu_2 \mu_3}
+ (k_3 - k_1)_{\mu_2} g_{\mu_3 \mu_1},   \\
\widehat{V}_{\mu_{1} \mu_{2} \mu_{3}}^{\kappa} (k_1, k_2, k_3) &=&
k_{3 \mu_1} g_{\mu_2 \mu_3} - k_{3 \mu_2} g_{\mu_1 \mu_3},
\\
\widehat{V}^{\lambda}_{\mu_1 \mu_2 \mu_3}(k_1, k_2, k_3) & = &
(k_1. k_2) (k_{3 \mu_1} g_{\mu_2 \mu_3} -  k_{3 \mu_2} g_{\mu_1
  \mu_3})
+ (k_2. k_3) (k_{1 \mu_2} g_{\mu_1 \mu_3} -  k_{1 \mu_3} g_{\mu_1
\mu_2}) \nonumber\\
& + & (k_3. k_1) (k_{2 \mu_3} g_{\mu_1 \mu_2} -  k_{2 \mu_1}
g_{\mu_2 \mu_3})
+ k_{1 \mu_3} k_{2 \mu_1} k_{3 \mu_2} - k_{1 \mu_2}
k_{2 \mu_3} k_{3 \mu_1},
\end{eqnarray}
cf.~\cite{BBP,BBK}.
Here $Q_{\gamma}(V)$ denotes the electric charge of the
vector leptoquark and the $Z$-coupling is
$Q_Z(V) = [T_3(V) - Q_{em}(V)
\sin^2 \theta_W]/(\cos \theta_W \sin \theta_W)$,
with $T_3$ being the third component of the
weak isospin and $\theta_W$ the electroweak mixing angle. In the
following we refer to the vector
leptoquarks discussed in ref.~\cite{BRW}.
Their quantum numbers can be found in ref.~\cite{BR},~table~1.

The differential scattering cross section in the center of momentum frame
reads
\begin{eqnarray}
\label{eqDIF}
\frac{d \sigma_{V\overline{V}}}{d \cos \theta} &=& \frac{3 \pi \alpha^2}
{2 M_V^2} \beta^3 \sum_{a = L,R}
\Biggl \{
A_a^{\gamma}~\widehat{T}_{\gamma}(\beta, \cos \theta,
\kappa_{\gamma},\lambda_{\gamma}) +
A_a^{\gamma Z}(s)
\widehat{T}_{\gamma Z}(\beta, \cos \theta, \kappa_{\gamma,Z},
\lambda_{\gamma,Z}) \nonumber\\ & &~~~~~~~~~~~~~~~~~~+
A_a^{Z}(s) \widehat{T}_{Z}(\beta, \cos \theta,
\kappa_Z,\lambda_Z) \Biggr \},
\end{eqnarray}
where $M_V$ denotes the leptoquark mass,
$\beta = \sqrt{1 - 4M_V^2/s}$, $s$ is
the cms energy squared,
\begin{eqnarray}
\label{eqA}
A_a^{\gamma}         &=& Q_{\gamma}^2(V) Q_{\gamma}^a,
  \\
A_a^{\gamma Z}(s)    &=&  2 Q_{\gamma}(V) Q_{\gamma}^a
Q_Z(V) Q_Z^a
                         \frac{(s - M_Z^2)s}
                         {(s - M_Z^2)^2 + \Gamma_Z^2 M_Z^2}, \\
A_a^{Z}(s)           &=& Q_Z^2(V) {Q_Z^a}^2
\frac{s^2}{(s - M_Z^2)^2 + \Gamma_Z^2 M_Z^2},
\end{eqnarray}
and
\begin{eqnarray}
\label{eqFH}
\widehat{T}_{\mu}(\beta, \cos \theta,
\kappa_{\mu},\lambda_{\mu})  &=& 1 - \kappa_{\mu} + \lambda_{\mu}
+ \frac{1}{4} (\kappa_{\mu} - \lambda_{\mu})^2
+ \frac{s}{16M_V^2} \sin^2 \theta \left [ (1-
\kappa_{\mu} )^2  + 2 \lambda_{\mu}^2
\right ] \nonumber\\
&+& \frac{1}{16} \sin^2 \theta  \left [ -1 -3 \beta^2 - 2( \kappa_{\mu}
- \lambda_{\mu})^2 + 4 \kappa_{\mu} \right ], \\
\widehat{T}_{\mu,\nu}(\beta, \cos \theta,
\kappa_{\mu,\nu},\lambda_{\mu,\nu})  &=&
1 - \frac{1}{2} \left(
\kappa_{\mu} + \kappa_{\nu} - \lambda_{\mu} - \lambda_{\nu} \right)
+ \frac{1}{4} \left ( \kappa_{\mu} - \lambda_{\nu})( \kappa_{\nu} -
\lambda_{\mu}
\right ) \nonumber \\
&+&  \frac{s}{16 M_V^2} \sin^2 \theta  \left [  (1 - \kappa_{\mu})
(1 - \kappa_{\nu}) + 2 \lambda_{\mu} \lambda_{\nu} \right]
\nonumber\\
&+& \frac{1}{16} \sin^2 \theta
\left [ - 1 - 3 \beta^2 + 2 (\kappa_{\mu} - \lambda_{\mu})
(\kappa_{\nu} -\lambda_{\nu})
+ 4 (\kappa_{\mu}
+    \kappa_{\nu} ) \right].
\end{eqnarray}
The lepton--gauge boson couplings are
$Q_{\gamma}^{L,R} = -1$,
$Q_Z^L = (-\frac{1}{2} + \sin^2 \theta_W)/\cos\theta_W \sin\theta_W$,
and
$Q_Z^R = \tan \theta_W$.

The angular distribution depends strongly on the value of the anomalous
couplings. For symmetric couplings $\kappa_{\gamma} = \kappa_Z$ and
$\lambda_{\gamma} = \lambda_Z$ the constant part vanishes for the case
$\kappa = \lambda + 2$. The differential cross section $d \sigma/d \cos
\theta$ is then
proportional to $\sin^2 \theta$ as in the case of scalar
leptoquark pair production~\cite{BR}. 
On the other hand, the differential distribution
turns out to be rather flat for minimal vector
couplings
$\kappa = 1, \lambda =0$,~cf.~ref.~\cite{BR}.

For the integrated cross section one obtains
\begin{eqnarray}
\label{eqSIG}
\sigma_{V \overline{V}}(s) &=&  \sigma_{\gamma} + \sigma_{\gamma,Z}
                           + \sigma_{Z} \\
&=&
\frac{3 \pi \alpha^2}{M_V^2} \beta^3 \sum_{a=L,R}
\left \{
A_a^{\gamma} T_{\gamma}(\beta,\kappa_{\gamma},\lambda_{\gamma}) +
A_a^{\gamma Z}(s) T_{\gamma Z}(\beta,\kappa_{\gamma,Z}
\lambda_{\gamma,Z}) +
A_a^{Z}(s) T_{Z}(\beta,\kappa_Z,\lambda_Z) \right \}. \nonumber
\end{eqnarray}
Here the functions $T^{\mu}(\beta,\kappa_{\mu},\lambda_{\mu})$ and
$T^{\mu\nu}(\beta,\kappa_{\mu,\nu},\lambda_{\mu,\nu})$ are given by
\begin{eqnarray}
\label{eqF}
T^{\mu}(\beta,\kappa_{\mu},\lambda_{\mu})
&=& \frac{(1 - \kappa_{\mu})^2
+ 2 \lambda_{\mu}^2}{24} \frac{s}{M_V^2} +
\frac{(23 - 20 \kappa_{\mu} + 4 \kappa^2_{\mu}) - 3 \beta^2}{24}
+ \lambda_{\mu} \left( 1 -
 \frac{\kappa_{\mu}}{3} + \frac{\lambda_{\mu}}{6} \right ),
\nonumber\\
T^{\mu\nu}(\beta,\kappa_{\mu,\nu},\lambda_{\mu,\nu}) &=&
\frac{(1 - \kappa_{\mu})(1 - \kappa_{\nu})
+ 2 \lambda_{\mu} \lambda_{\nu}}{24} \frac{s}{M_V^2} +
\frac{[23 - 10 (\kappa_{\mu} + \kappa_{\nu})
+ 4 \kappa_{\mu} \kappa_{\nu}]
- 3 \beta^2}{24} \nonumber\\ &+&
  \frac{\lambda_{\mu}}{2}
\left( 1 - \frac{\kappa_{\nu}}{3} + \frac{\lambda_{\nu}}{6} \right )
+ \frac{\lambda_{\nu}}{2}
\left( 1 - \frac{\kappa_{\mu}}{3} + \frac{\lambda_{\mu}}{6} \right ).
\end{eqnarray}

The minimal cross section among all possible choices of the four
anomalous couplings
$\kappa_{\gamma}, \kappa_Z, \lambda_{\gamma}$ and $\lambda_Z$,
is
\begin{equation}
\label{eqSIGMIN}
\sigma_{V \overline{V}}^{min}(s, M_V^2)
= \frac{\pi \alpha^2}{2 M_V^2} \beta^7 \sum_{a=L,R}
|\ka(s)|^2 \frac{1}{3(5 - 3 \beta^2)},
\end{equation}
where
\begin{equation}
\label{eqKA}
\ka(s) = \sum_{B = \gamma, Z} Q_B^a \frac{s}
{s - M_B^2 + i M_B \Gamma_B} Q_B(V).
\end{equation}
Unlike in
the case of general values of the anomalous couplings $\kappa$
and $\lambda$~(\ref{eqSIG})
the minimal cross section does {\it not}
 contain unitarity violating
contributions $\propto s/M_V^2$.
The anomalous couplings determining the minimal cross
section~(\ref{eqSIGMIN})
are depicted in figure~1. Due to the symmetry of the quadratic form
(\ref{eqSIG}) they are given by
\begin{eqnarray}
\label{eqANOeq1}
\kappa^{min}_{\gamma} =
  \kappa^{min}_Z
 = \frac{1}{2} \left  (
\frac{15 - 10 \beta^2 - \beta^4}{5 - 3 \beta^2}    \right ),
\\
\label{eqANOeq2}
\lambda^{min}_{\gamma} =
\lambda^{min}_Z
 = -\frac{1}{2}
(1 - \beta^2)  \left ( \frac{5 - \beta^2}
{5 - 3 \beta^2} \right ).
\end{eqnarray}
Only for large cms energies, $\beta \rightarrow 1$, ($S \gg 4 M_V^2$),
the minimizing anomalous couplings approach
the minimal vector couplings $\kappa^{min} = 1$ and $\lambda^{min} = 0$.
On the other hand, for  $\beta \rightarrow 0$ the couplings
$\kappa^{min} = 3/2$ and $\lambda^{min} = -1/2$ are obtained.

If one would assume $\lambda_{\gamma} \equiv \lambda_{Z} \equiv 0$
and minimize the cross section for $\kappa_{\gamma}$ and $\kappa_Z$ 
only
\begin{equation}
\label{eqSIGMI1}
\sigma_{V \overline{V}}^{min}(s, M_V^2)
= \frac{\pi \alpha^2}{2 M_V^2} \beta^3 \sum_{a=L,R}
|\ka(s)|^2 \frac{10 - 13 \beta^2 + 10 \beta^4 - 3 \beta^6}
{24(2 - \beta^2)^2}
\end{equation}
is obtained, where
\begin{equation}
\label{eqKA1}
\kappa^{min}_{\gamma, Z} (\lambda_{\gamma, Z} \equiv 0) =
\frac{7 - 5 \beta^2}{2(2 - \beta^2)},
\end{equation}
i.e. $\kappa^{min} = 7/4$ for $\beta \rightarrow 0$.
Here the cross section grows $\propto \beta^3$ near
the threshold as also in the case of minimal vector couplings and 
Yang--Mills type couplings.
The weaker rise $\propto \beta^7$ in the previous case is
thus a consequence of considering  non--zero values of
$\lambda_{\gamma,Z}$ in eq.~(\ref{eqVERT}).

In figure~2 the mass dependence of the pair production cross section of
the leptoquark $U_1$~\footnote{
See refs.~\cite{BRW, BR} for the notation.} is shown at
$\sqrt{s} = 1 \TeV$ for different choices of the anomalous couplings,
upon which the accessible mass bounds  strongly depend.
Assuming a signal of 100 events at an integrated luminosity of
${\cal L} = 10~{\rm fb}^{-1}$ for the minimizing anomalous couplings,
a mass bound of $370 \GeV$ can be reached, whereas the corresponding
bound for minimal vector couplings is $470 \GeV$. A still larger
cross section  is obtained for Yang--Mills type couplings
$\kappa_{\gamma,Z} = \lambda_{\gamma,Z} \equiv 0$.

In figures~3a--c the minimal values of the integrated cross sections
are shown for all
leptoquark states of ref.~\cite{BRW} at $\sqrt{s} = M_Z,~190~\GeV$,
and $1~\TeV$. Accessible mass ranges are indicated assuming
100 signal events\footnote{The corresponding numbers are meant to be
indicative and can not replace a detailed analysis
accounting for all experimentally relevant aspects.}
at integrated luminosities of
${\cal L}(M_Z) = 150~{\rm pb}^{-1},~{\cal L}(190 \GeV) 
= 500~{\rm pb}^{-1}$, and
${\cal L}(1 \TeV) = 10~{\rm fb}^{-1}$, respectively.
For all cms energies considered,  the  lowest mass
bounds are estimated for
the states $U_1$ and $U_3(0)$, respectively, which have the same
bosonic quantum numbers. The largest cross sections are obtained for
$U_3(-1)$ for $\sqrt{s} = M_Z$ and $U_3(1)$ at $\sqrt{s} = 190~\GeV$
and $1~\TeV$, respectively.

In conclusion, we have shown that non--vanishing anomalous couplings
$\lambda_{\gamma}$ and $\lambda_Z$ may yield much weaker search limits
in the threshold range as obtained varying the couplings 
$\kappa_{\gamma}$
and $\kappa_Z$ {\it only}. This should be taken into account in
forthcoming analyses of the LEP data~\footnote{Previous
analyses~\cite{LEPEX} considered  only scalar leptoquarks.} both
for $\sqrt{s} = M_Z$, the energy range at LEP2, and at future linear
colliders.

\vspace{1mm}
\noindent
{\bf Acknowledgments:}~We
would like to thank Prof. P. S\"oding for his constant support of the
present project. We would like to thank S. Riemersma for   reading
the manuscript. E.B. and A.K. would like to thank DESY--Zeuthen for the 
warm hospitality extended to them.
The work has been supported in part by the EC grant `Capital Humain et
Mobilite' CHRX--CT92--0004,
by the grant 95-0-6.4-38 of the Center of Natural
Sciences of the State Committee for Higher Education in Russia,
and by the RFBR grants 96-02-18635a and 96-02-19773a.


\newpage
\begin{center}

\mbox{\epsfig{file=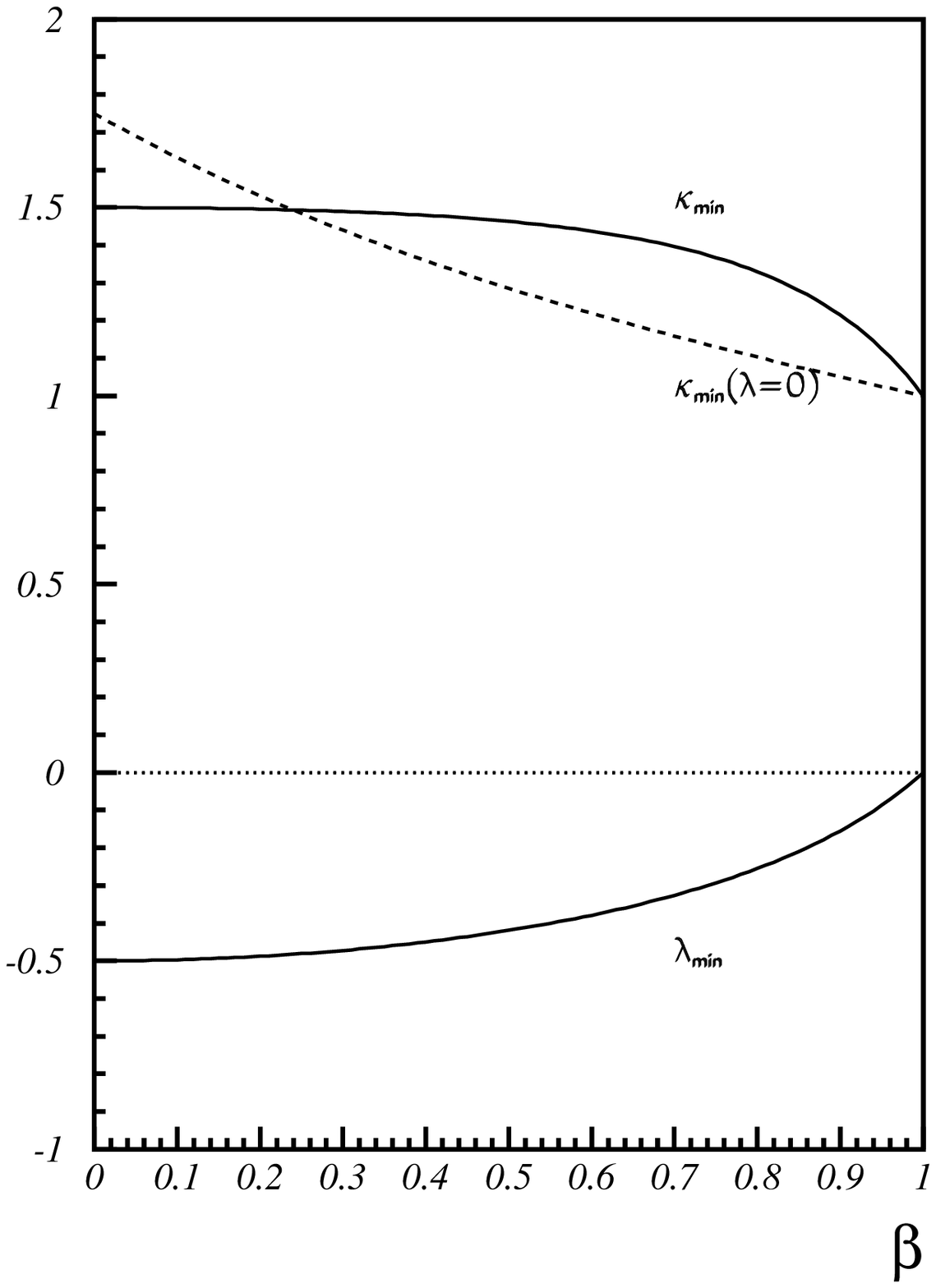,height=18cm,width=16cm}}

\vspace{2mm}
\noindent
\small
\end{center}
{\sf
Figure~1:~The minimizing couplings $\kappa^{min}$ and $\lambda^{min}$
as a function of $\beta = \sqrt{1 - 4 M_V^2/s}$. The dashed line
corresponds to the value of $\kappa$ minimizing the production cross
section for $\lambda = 0$.}
\normalsize
\newpage
\begin{center}

\mbox{\epsfig{file=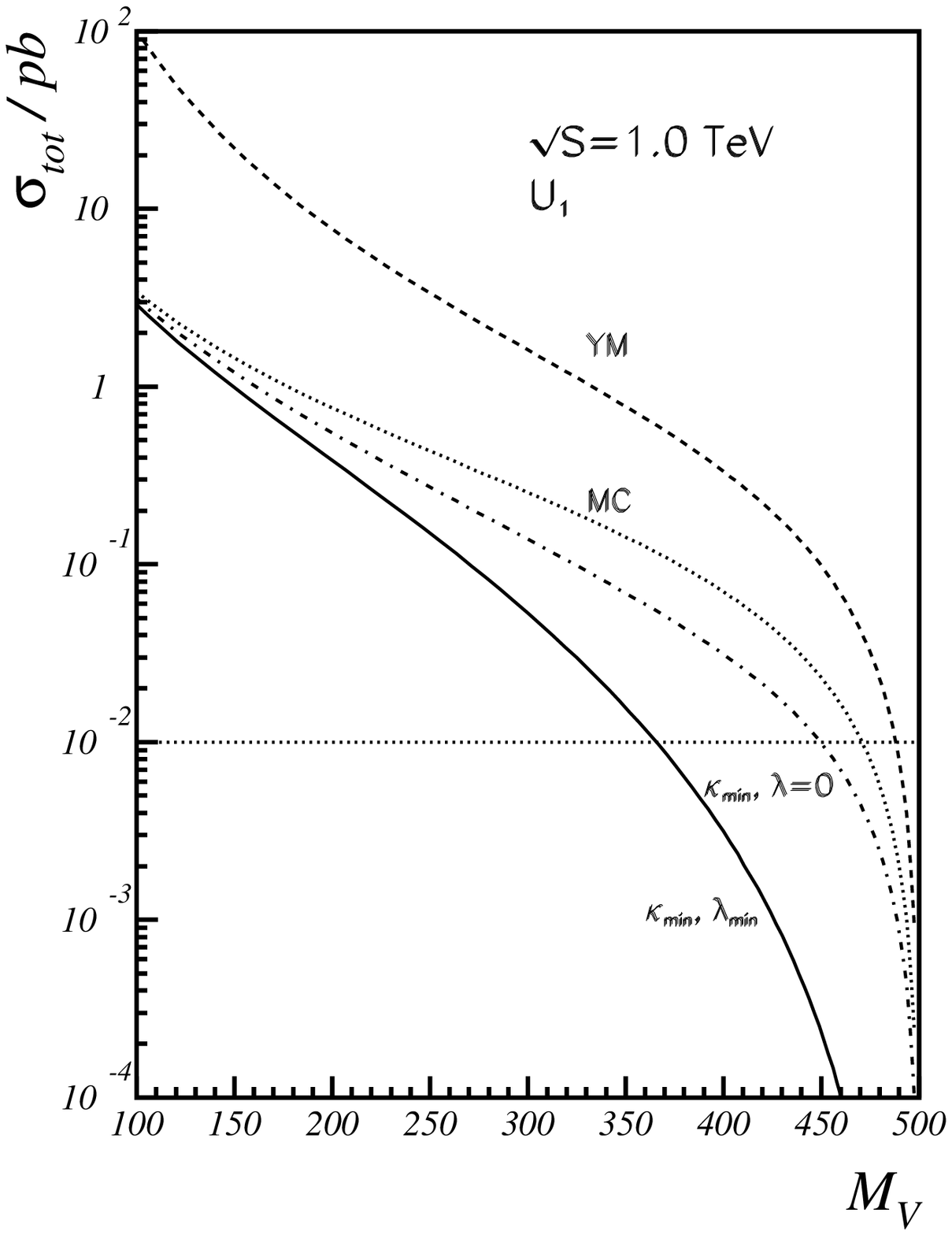,height=18cm,width=16cm}}

\vspace{2mm}
\noindent
\small
\end{center}
{\sf
Figure~2:~Integrated pair production cross section for the
vector leptoquark $U_1$, ref.~\cite{BRW},
 for different values of the anomalous couplings
$\kappa_{\gamma,Z}$ and $\lambda_{\gamma,Z}$. Solid line: minimal cross
section; dash--dotted line: minimal cross section for $\lambda_{\gamma}
= \lambda_Z =0$; dotted line: cross section for minimal vector
couplings (MC)~$\kappa_{\gamma,Z} = 1 , \lambda_{\gamma,Z}= 0$; dashed
line: cross section for Yang--Mills type couplings 
(YM)~$\kappa_{\gamma,Z} = \lambda_{\gamma,Z}= 0$. 
The horizontal dotted
line indicates the accessible search range at a luminosity of ${\cal L}
= 10~{\rm fb}^{-1}$ and 100 signal events.
\normalsize
\newpage
\begin{center}

\mbox{\epsfig{file=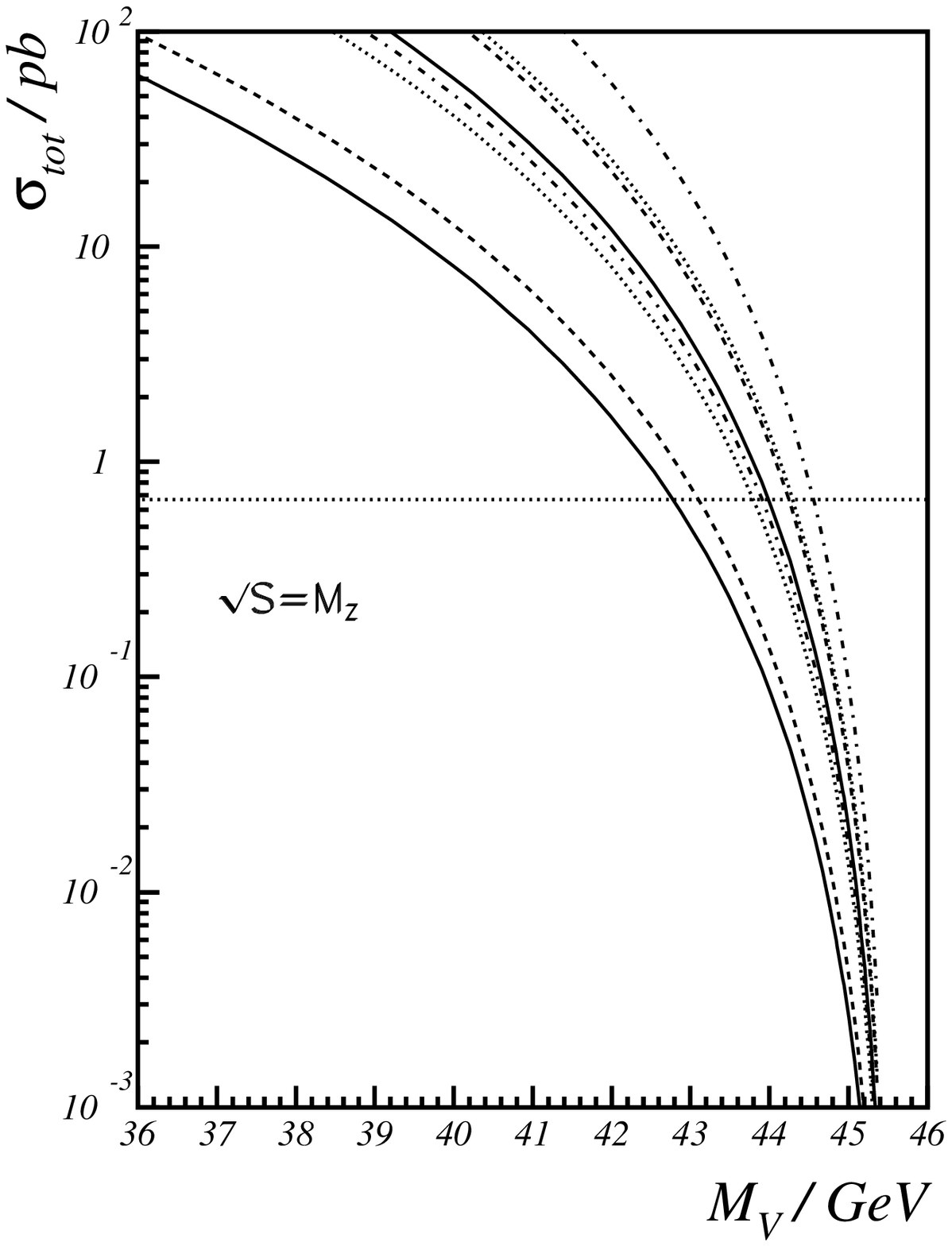,height=18cm,width=16cm}}

\vspace{2mm}
\noindent
\small
\end{center}
{\sf
Figure~3a:~Mass dependence of the minimal integrated pair production
cross section for the vector leptoquarks ref.~\cite{BRW} at
$\sqrt{s} = M_Z$. The lines correspond to the production cross
sections for the states $U_1~(U_3(0)), V_2(1/2), \widetilde{V}_2(-1/2),
\widetilde{U}_1, \widetilde{V}_2(1/2), V_2(-1/2), U_3(1)$ and $U_3(-1)$
from left to  right. The horizontal line
indicates the accessible search range at a luminosity of ${\cal L}
= 150~{\rm pb}^{-1}$ and 100 signal events.
}
\normalsize
\newpage
\begin{center}

\mbox{\epsfig{file=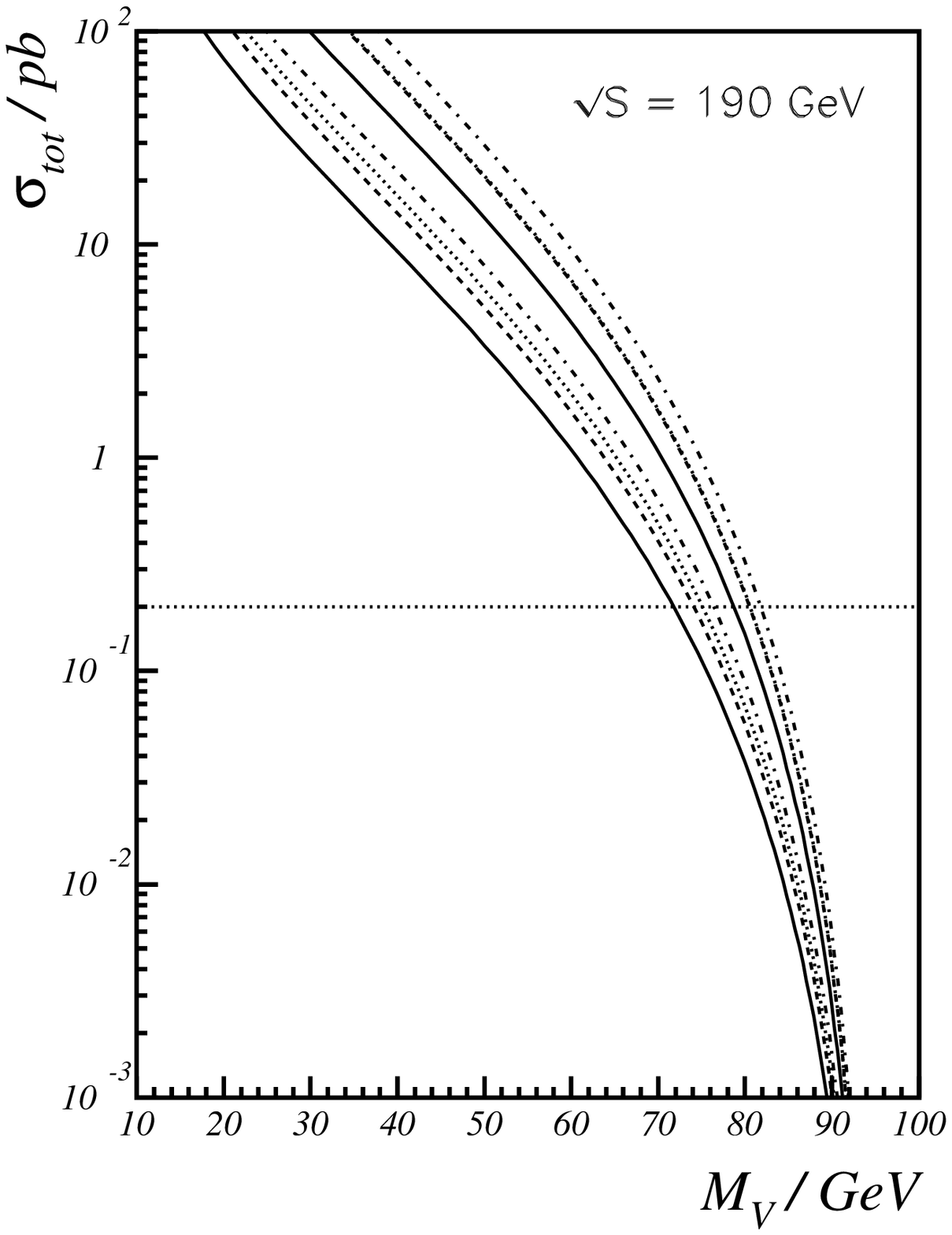,height=18cm,width=16cm}}

\vspace{2mm}
\noindent
\small
\end{center}
{\sf
Figure~3b:~Mass dependence of the minimal integrated pair production
cross section for the vector leptoquarks ref.~\cite{BRW} at
$\sqrt{s} = 190 \GeV$. The lines correspond to the production cross
sections for the states $U_1~(U_3(0)),  \widetilde{V}_2(1/2),
\widetilde{V}_2(-1/2), V_2(-1/2), V_2(1/2), U_3(-1)
(\widetilde{U}_1)$ and  $U_3(1)$
from left to  right. The horizontal line
indicates the accessible search range at a luminosity of ${\cal L}
= 500~{\rm pb}^{-1}$ and 100 signal events.
}
\normalsize
\newpage
\begin{center}

\mbox{\epsfig{file=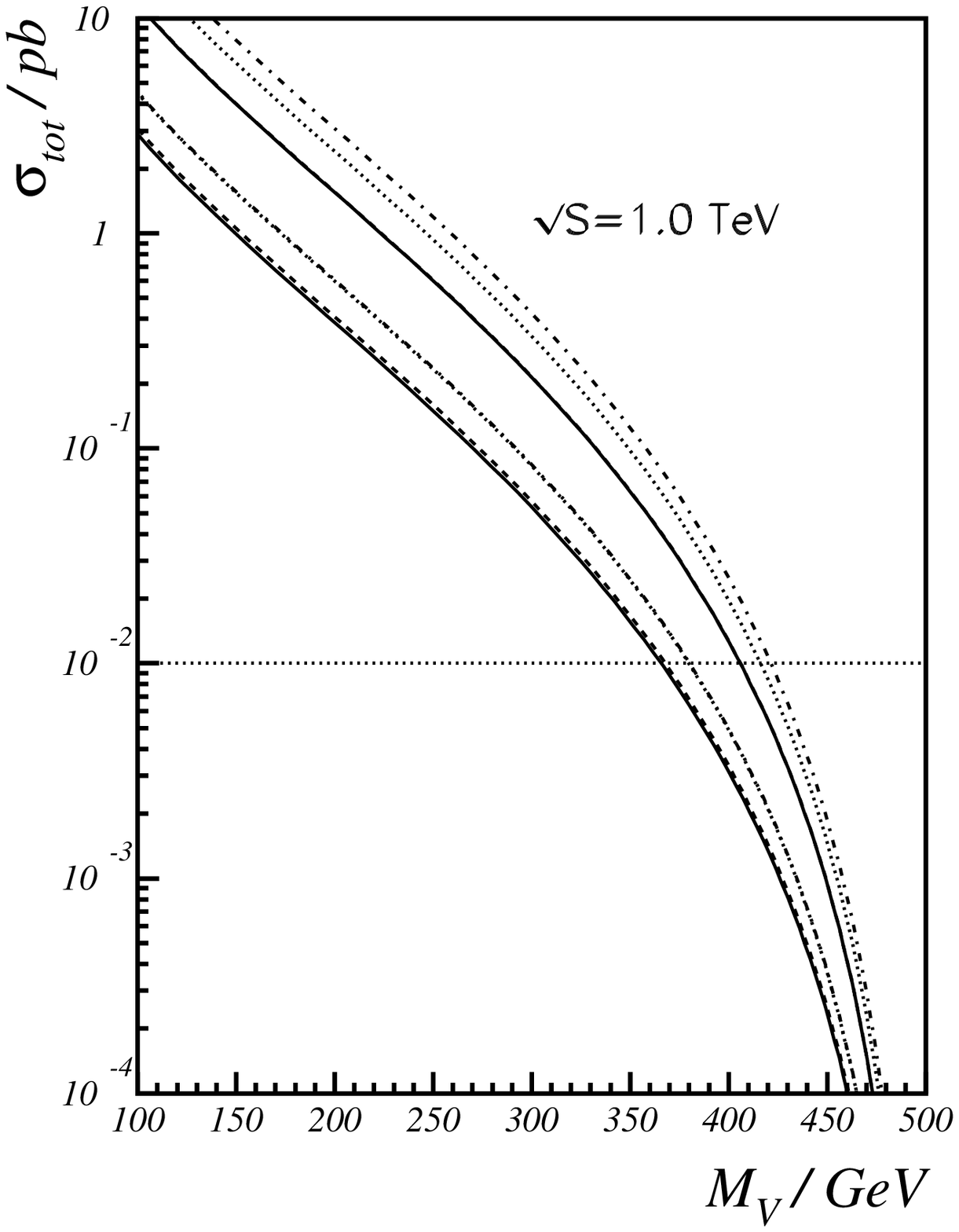,height=18cm,width=16cm}}

\vspace{2mm}
\noindent
\small
\end{center}
{\sf
Figure~3c:~Mass dependence of the minimal integrated pair production
cross section for the vector leptoquarks ref.~\cite{BRW} at
$\sqrt{s} = 1~\TeV$. The lines correspond to the production cross
sections for the states $U_1~(U_3(0)),  \widetilde{V}_2(1/2),
\widetilde{V}_2(-1/2)(V_2(-1/2)),  U_3(-1)(V_2(1/2)),
\widetilde{U}_1$ and  $U_3(1)$
from left to  right. The horizontal line
indicates the accessible search range at a luminosity of ${\cal L}
= 10~{\rm fb}^{-1}$ and 100 signal events.
}
\normalsize

\begin{thebibliography}{999}
%
\bibitem{BRW}
W. Buchm\"uller, R. R\"uckl, and D. Wyler, Phys. Lett.
{\bf B191} (1987) 442.
%
\bibitem{LEUR}
W. Buchm\"uller and D. Wyler, Phys. Lett. {\bf B177} (1986) 377;\\
M. Leurer, Phys. Rev. Lett. {\bf 71} (1993) 1324; Phys. Rev. {\bf D49}
(1994) 333; Phys. Rev. {\bf D50} (1994) 536;\\
S. Davidson, D. Bailey, and B.A. Campbell, Z. Phys. {\bf C61} (1994) 613.
%
\bibitem{BR}
J. Bl\"umlein and R. R\"uckl, Phys. Lett. {\bf B304} (1993) 337.
%
\bibitem{BD}
J.D. Bjorken and S.D. Drell, {\sf Relativistic Quantum Mechanics},
(McGraw--Hill, New York, 1964).
%
\bibitem{KALAX}
W. Buchm\"uller and D. Wyler, Nucl. Phys. {\bf B268} (1986) 621;\\
K. Hagiwara, S. Ishihara, R. Szalapski, and D. Zeppenfeld, Phys. Rev.
{\bf D48} (1993) 2182, and references therein.
%
\bibitem{BB}
J. Bl\"umlein and E. Boos, Nucl. Phys. {\bf B} (Proc. Suppl.),
{\bf 37B} (1994) 181.
%
\bibitem{BBK}
J. Bl\"umlein, E. Boos, and A. Kryukov, DESY 96--174,
{\tt hep-ph/9610408}.
%
\bibitem{BBP}
J. Bl\"umlein, E. Boos, and A. Pukhov, Mod. Phys. Lett. {\bf A9} (1994)
3007.
%
\bibitem{LEPEX}
D. Alexander et al., OPAL collaboration, Phys. Lett. {\bf B275} (1992)
123; Phys. Lett. {\bf B263} (1991) 123;\\
D. Decamp et al., ALEPH collaboration, Phys. Rep. {\bf 216} (1992) 253;\\
B. Adeva et al., L3 collaboration, Phys. Lett. {\bf B261} (1991) 169;
O. Adriani, L3 collaboration, Phys. Rep. {\bf 236} (1993) 1; \\
P. Abreu et al.,  DELPHI collaboration, Phys. Lett. {\bf B275} (1992) 222;
{\bf B316} (1993) 620.
\normalsize
\end{thebibliography}
\end{document}